\documentclass[a4paper,man,natbib]{apa6}

\usepackage[english]{babel}
\usepackage[utf8x]{inputenc}
\usepackage{amsmath}
\usepackage{graphicx}
\usepackage[colorinlistoftodos]{todonotes}
\usepackage{bm}

\newcommand{\y}{\mbox{$\mathbf{y}$}}
\newcommand{\ys}{\mbox{$\bm{y}_s$}}

\newcommand{\yrepk}{\mbox{$\y_{\text{rep}}^{(k)}$}}

\newcommand{\ns}{\mbox{$n_s$}}

\newcommand{\condrespone}{\mbox{$\pi_1c$}}

\newcommand{\classp}{\mbox{$\rho_c$}}

\newcommand{\Drisk}{\mbox{$D_{\text{risk}}(\y, R)$}}
\newcommand{\Drisksix}{\mbox{$D_{\text{risk}}(\y, 6)$}}

\title{The Lazy Bootstrap. A Fast Resampling Method for Evaluating Latent Class Model Fit}
\shorttitle{The Lazy Bootstrap}
\author{\mbox{Geert H. van Kollenburg$^1$}, \mbox{Joris Mulder$^2$}, \mbox{Jeroen K. Vermunt$^2$} }
\affiliation{ \mbox{$^1$Radboud University, Nijmegen, Netherlands} \mbox{$^2$Tilburg University, Tilburg, Netherlands}}

\abstract{The latent class model is a powerful unsupervised clustering algorithm for categorical data. Many statistics exist to test the fit of the latent class model. However, traditional methods to evaluate those fit statistics are not always useful. Asymptotic distributions are not always known, and empirical reference distributions can be very time consuming to obtain. In this paper we propose a fast resampling scheme with which any type of model fit can be assessed. We illustrate it here on the latent class model, but the methodology can be applied in any situation. 
	
	The principle behind the lazy bootstrap method is to specify a statistic which captures the characteristics of the data that a model should capture correctly. If those characteristics in the observed data and in model-generated data are very different we can assume that the model could not have produced the observed data. With this method we achieve the flexibility of tests from the Bayesian framework, while only needing maximum likelihood estimates. We provide a step-wise algorithm with which the fit of a model can be assessed based on the characteristics we as researcher find important. In a Monte Carlo study we show that the method has very low type I errors, for all illustrated statistics. Power to reject a model depended largely on the type of statistic that was used and on sample size. We applied the method to an empirical data set on clinical subgroups with risk of Myocardial infarction and compared the results directly to the parametric bootstrap. The results of our method were highly similar to those obtained by the parametric bootstrap, while the required computations differed three orders of magnitude in favour of our method.}
\keywords{Goodness-of-fit, p-values, latent class analysis, validation, bootstrap}

\begin{document}
\maketitle
The latent class (LC) model \citep{goodman1974exploratory} is a powerful unsupervised clustering algorithm for categorical data that is currently being used in a wide range of research fields. An important part of doing LC analysis is to assess the fit of the model to the observed data. Besides the traditional $\chi$-squared goodness-of-fit statistics, various specific statistics have recently been developed for this model and its extensions \citep{vermunt2016technical,oberski2013monte,nagelkerke2016BVR,nagelkerke2016power}. 

Deciding on whether the value of a statistic indicates model misfit is usually based on a $p$ value, which quantifies how likely the observed data are if the employed model holds in the population. The asymptotic $p$ value is the most commonly used $p$ value. It is very easy to calculate, but can only be obtained if the asymptotic distribution of a statistic is known, which is not the case for all available statistics \citep{kollenburg2015assessing,oberski2013monte}. Additionally, the asymptotic approximation may break down when sample sizes are not large so that the resulting $p$ values can become very unreliable \citep{kollenburg2015assessing}. More reliable $p$ values can be obtained by using computational intensive resampling schemes yielding empirical reference distributions. This is most commonly done by means of the parametric bootstrap \citep{efron1993introduction}. The downside of the bootstrap method is that it can become very time-consuming when the models under investigation are complex, such as the LC model, because the same model has to be estimated many times \citep{nagelkerke2016power}. And even then, certain statistics may still result in unreliable $p$ values \citep{von1997bootstrapping,maydeu2006limited,kollenburg2015assessing,oberski2013monte}; for example, bootstrap $p$ values for overall goodness-of-fit statistics are unreliable when the number of possible data patterns is very large. Efforts to speed up the bootstrap procedure usually focus on parameter estimation \citep{antal2014new,salibian2002bootstrapping} or are limited to small data sets \citep{kisielinska2013exact}. In the Bayesian framework, there is a tool available to test model fit, called the posterior predictive check \citep{meng1994posterior,gelman1996posterior}. Though it is fast and does not require repeated model estimation, the need for MCMC algorithms and choices for prior distributions may be a barrier for applied researchers.

Ideally, a test for model fit should be fast, reliable, and easy to implement. The methodology proposed in this paper improves on existing methods by eliminating the need for repeated model estimation, while still only requiring maximum likelihood estimates. The idea is to directly compare observed data with model-generated data in a way that requires no time-consuming model estimation procedures. This comparison can be based on characteristics corresponding to those aspects of the data that the researcher finds most important. If the observed data differ consistently from the model-generated data, we have strong evidence that the model under consideration could not have produced the observed data. The proposed methodology cannot only be used to assess how well a model can explain particular aspects of the data, it can also be used to assess whether the key model assumptions are met when applying the model concerned on the data set at hand. As we will show, the proposed methodology provides great flexibility to assess many different aspects of model fit. 

As an example, suppose we are interested in testing whether a latent LC model is appropriate to describe a data set based on a risk inventory. For each observation, a number of variables are scored according to whether a particular risk is present (scored 1) or not present (scored 0). A first research question would be how many latent subgroups are needed to capture the associations between those variables. Second, we may require the model to correctly capture the observed frequency of observations who have all risk present. For the first research question, it is straightforward to quantify the association by a statistic such as the Pearson's $\chi^2$ statistic under independence. For the second question, the statistic of interest is simply the observed frequency that this particular pattern occurred. Suppose we have an observed $\chi^2=200$ and that, out of 100 observations, the pattern with all risks present occurs 25 times. We then generate many data sets from the estimated latent class model. If in those generated data sets we find $\chi^2$-values around 200 and frequencies of the response pattern of interest of around 25, the employed model seems valid to explain the characteristics of the observed data. The reason is that the data generated from the fitted LC model has similar characteristics as the observed data. 

In this paper, we shall focus on tests to assess whether a LC model adequately captures specific characteristics of the data. We will achieve this by using test statistics which capture characteristics of a data set and which can be calculated directly from the data (e.g., descriptive statistics) without needing to re-estimate the model for each replicated data set using time-consuming algorithms such as the Expectation Maximization (EM) algorithm. Although the proposed procedure is similar to the parametric bootstrap \citep{efron1993introduction}, the key difference is that our methodology directly compares observed and replicated data, rather than in an indirect manner as is done when using a bootstrap of traditional goodness-of-fit statistics. It should be noted that the parametric bootstrap needs to estimate the model of interest on both the observed data and the (large number of) generated data sets because a goodness-of-fit statistic needs model estimates for its calculation. Due to the indirect nature of testing with the parametric bootstrap, our method has the advantage that it can be up to hundreds of times faster, because the model only needs to be estimated once rather than hundreds of times as is required for the parametric bootstrap. The second advantage of using tailored test statistics is that we obtain a detailed picture of which aspects of the data are and which aspects are not captured by the model. In contrast to the posterior predictive check \citep{meng1994posterior,gelman1996posterior}, the proposed method is based on maximum likelihood estimation, which is generally easier to perform than Bayesian MCMC sampling and does not require specification of prior distributions.

In the remainder of this paper we will illustrate the proposed resampling methodology within the context of LC analysis. In the next section we discuss the LC model in detail and provide statistics with which the fit of a LC model can be tested. After that we introduce the new methodology in more detail, and provide a step-wise algorithm for applying the method to LC analysis. We include a simulation study evaluating type I error rates and power for various statistics of interest. The paper ends with a discussion on the main finding, their implications for applied researchers, and interesting topics for future research.

\section{The Latent Class Model}

The LC model \citep{goodman1974exploratory} is used to cluster observations when there is no knowledge about which observation belongs to which group (i.e., it is unsupervised). Suppose we have $N$ observations on $J$ dichotomous variables (e.g., yes/no, present/absent, male/female). Each observation can have one of $S = 2^J $ possible response patterns $\ys = (y_{s1}, \ldots, y_{sJ})$, for $s = 1, \ldots, S$. The LC model assumes that the $N$ observations can be divided into $C$ distinct classes/categories of an unobserved variable based on their response pattern. 

According to the LC model, each class has a different probability of giving a particular response to the variables. We denote the probability of giving response 1 to variable $j$ given that an observation belongs to class $c$ as $\pi_{jc}$. The (unknown) class sizes are denoted by $\rho_c$. The LC model further assumes that within a class, the responses to the variables are independent (i.e., local independence). Therefore, we can write the LC model likelihood of observing a particular response pattern as 
\begin{equation} \label{eq:LC}
Pr(\ys)=\sum^{C}_{c=1} \rho_c \prod_{j=1}^{J}\pi_{jc}^{y_{sj}}(1-\pi_{jc})^{1-y_{sj}}.
\end{equation}
As an example, the response probability for a pattern with only ones on all $J$ variables is equal to $\sum_{c=1}^C \rho_c \times \pi_{1c}\times\ldots\times \pi_{Jc}$. Maximum likelihood estimates for the model parameters are usually obtained though the Expectation-Maximization (EM) algorithm (Dempster, Laird, \& Rubin, 1977). 

\subsection{Statistics for the latent class model.}

The focus of LC analysis often lies with explaining associations between variables. A straightforward choice to assess the strength of the association between categorical variables is to calculate a chi-squared type of statistic \citep{agresti1987empirical} quantifying the deviation from independence. Commonly used measures are Pearson's $X^2$,
\[
X^2 = \sum_{s=1}^{S}{\frac{(n_{s}-e_{s})^2}{e_{s}}},
\]
and the likelihood ratio chi-square,
\[
G^2 = 2\sum_{s=1}^{S} n_s \ln (n_s/e_s), 
\]
where $n_s$ is the observed frequency of a particular pattern $\ys$ and $e_s$ is the expected frequency of that pattern under independence. It is important to note that the $ e_s$ can easily be computed from the data, that is, by multiplying the univariate marginal frequencies of the variables involved. These chi-squared type of statistics can be used to quantify associations of any order, from second-order for bivariate associations to $J$-th order for the overall association between all $J$ variables used in the analysis. 

Another example of a question that a researcher may be interested in is whether we can find an explanation for the occurrence of particular response patterns. For instance, suppose we have data on some sort of risk inventory. Each dichotomous variable in the inventory indicates whether a particular risk is present (scored 1) or not present (scored 0). Now suppose that we focus on the number of high-risk observations for which at least a certain number (say, $Q$) of the risks are present. The statistic used to test this may simply be the total frequency of the patterns with $Q$ or more risks present: 
\[
\Drisk = \sum_{s=1}^S \ns * I(\sum_{j=1}^J y_{sj} \geq Q)  
\]
where $\ns$ is the observed frequency of pattern $\ys$ and where the indicator function $I$ equals 1 if the sum of the scores in pattern $s$ is greater than or equal to $Q$ (i.e., it has $Q$  or more risks present) and 0 otherwise. The formula then sums all the corresponding frequencies, $\ns$. Aside from risk assessment, other examples include presence of symptoms, giving correct answers, and agreement to statements. The statistic can easily be adapted to assess whether the frequency of one particular response pattern (e.g., $[1,0,0,1,0,1]$) is correctly picked up. In that case, the statistic is reduced to simply the observed frequency $n_s$ for the pattern of interest.

The LC model is of particular interest for the above stated research questions, because it may explain that the associations found in the data are a result of the fact that the data is a combination of data from different subgroups. Additionally, the LC model might explain that the observed frequency of risk patterns is due to the fact that there is a specific subgroup which is prone to have high risks, while another subgroup has not.

It is important to note that the chi-squared statistics that we use in the current application are traditionally used within LC modelling as residuals to test \emph{remaining} association given a particular model. The fit of a LC model is then usually evaluated through the parametric bootstrap \citep{oberski2013monte,kollenburg2015assessing,nagelkerke2016power}. When used as residuals, the expected frequencies $e_s$ are based on the ML estimates (plugged into Equation~\ref{eq:LC}). The implication is that ML estimates also have to be found for every replicate data set, which due to the need for iterative estimation algorithms may make the parametric bootstrap very time consuming. The field can therefore greatly benefit from the use of tailor-made statistics which can be calculated directly on the data.

\section{A New Methodology to Test Model Fit}

The general idea of resampling methods is to obtain an approximation of the distribution of a statistic without the requirement of relying on asymptotics. To assess model fit, we check whether a model of interest has a similar fit to the observed data (quantified by a goodness-of-fit statistic) as it has to (hypothetical) model-generated data (quantified by the same statistic). When the fit in replicated data and observed data are similar, this suggests that the model fits the data well. When the fit in the observed data is much worse than in the generated data, this suggests that the model does not fit. The underlying principle is that if the observed data was generated by the model, it should be similar to artificial data of which we know that it was generated from the model. 

The methodology proposed here applies the same basic principle of all resampling methods, that is, by quantifying important characteristics of the data and comparing those characteristics in observed and model-generated data. The following algorithm describes how to assess the fit of a LC model based on a statistic calculated directly on the data. The algorithm can be applied to any statistic and any model specification. For illustration purposes we will include in the description how to use the $X^2$ for testing the fit of a 2-class model.  \\
\bigskip

\begin{figure}[t]
	\centering
	\makebox{\includegraphics[width=14.0cm, height = 10cm]{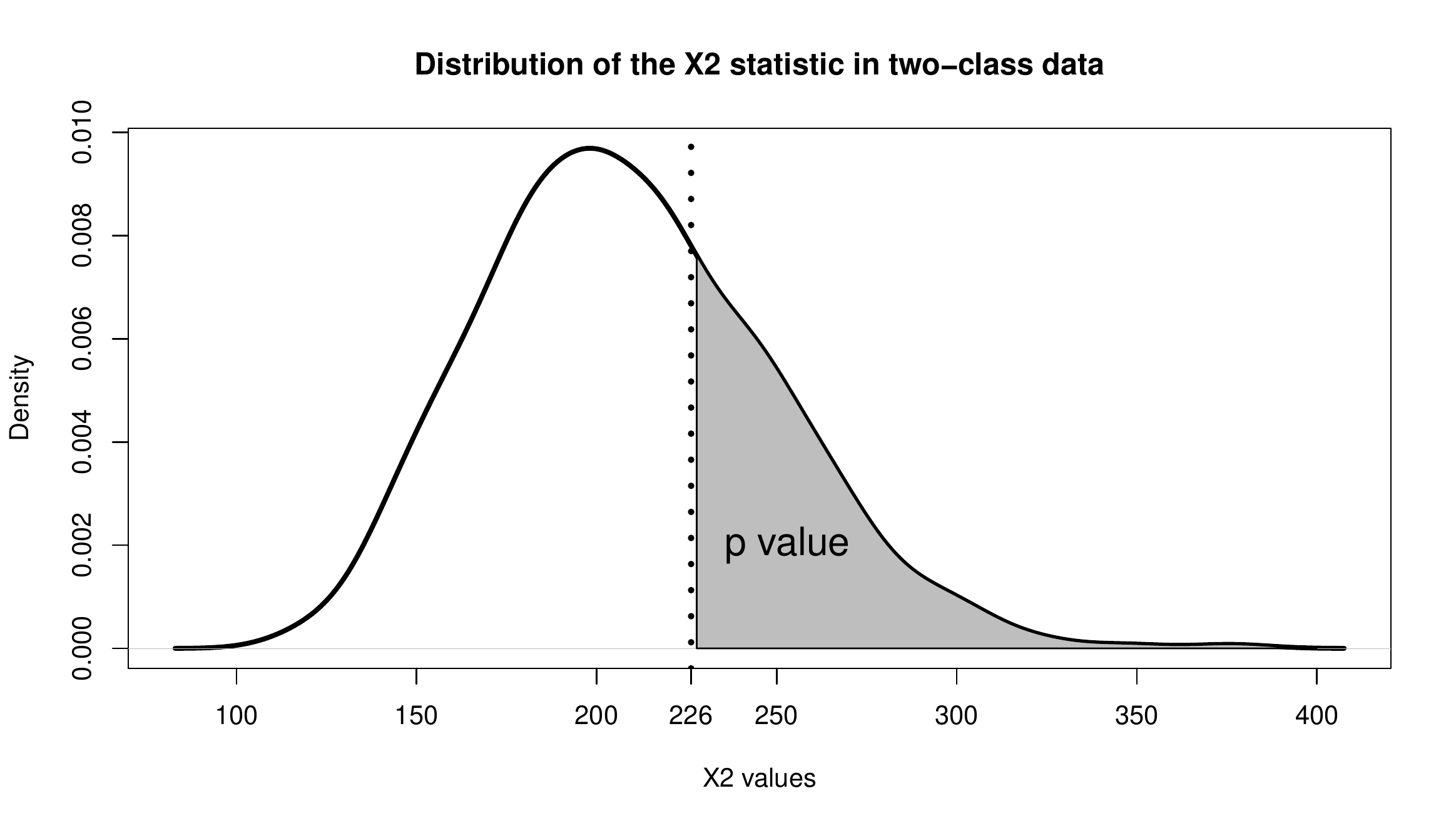}}
	\caption{The distribution of the $X^2$ statistic calculated on 500 data sets generated from a LC model with 2 classes. The value of $X^2$ in the observed data is represented by a vertical dotted line. The proportion of the area under the curve to the right of the vertical line (i.e., .296) is used to make the decision about the appropriateness of the model.}
	\label{fig:example}
\end{figure}

\noindent\textbf{Algorithm 1: Comparing characteristics of observed and model-generated data}
\begin{description}
	\item[Step 1:] Specify the important characteristics of the observed data and calculate the corresponding statistic. 
	
	We are interested in correctly reproducing the overall association between the variables as quantified by 
	\[
	X^2 =  \sum_{s=1}^{S}{\frac{(n^{\text{obs}}_{s}-e^{\text{obs}}_{s})^2}{e^{\text{obs}}_{s}}}.
	\]
	
	\item[Step 2:] Specify the likelihood of the model and obtain ML estimates for the model parameters.

	The likelihood for a particular response in a 2-class model is given by
	\begin{equation}
	Pr(\ys)=\sum^{2}_{c=1} \rho_c \prod_{j=1}^{J}\pi_{jc}^{y_{sj}}(1-\pi_{jc})^{1-y_{sj}},
	\end{equation}
	where ML estimates for the parameters can be obtained through the EM algorithm.

	\item[Step 3:] Obtain values for the chosen statistics in replicated data sets:
	\begin{description}
		\item[Step 3a:] Plug in the ML estimates into the likelihood specified in Step 2. 
		
		We calculate the probability for each response pattern, $\ys$, by plugging the ML estimates into the likelihood:
		\[
		\widehat{Pr}(\ys)=\hat{\rho}_1 \prod_{j=1}^{J}\hat{\pi}_{j1}^{y_{sj}}(1-\hat{\pi}_{j1})^{1-y_{sj}} + \hat{\rho}_2 \prod_{j=1}^{J}\hat{\pi}_{j2}^{y_{sj}}(1-\hat{\pi}_{j2})^{1-y_{sj}}.
		\]

		\item[Step 3b:] Generate a replicated data set, $\yrepk$ from the likelihood with the ML estimates plugged in.
		
		A replicate data set is generated by taking a draw from a multinomial distribution, with the estimated pattern probabilities as parameters:
		\[
		\yrepk \sim \text{Multinomial}(N|\widehat{Pr}(\y_1), \ldots \widehat{Pr}(\y_S)). 
		\]
		
		\item[Step 3c:] Calculate the chosen statistic on the replicated data set.

		We compute the association between all the variables in a replicate data set $\yrepk$ as: 
		\[
		X^{2, (k)} =  	\sum_{s=1}^{S}{\frac{(n^{\text{(k)}}_{s}-e_{s})^2}{e^{\text{(k)}}_{s}}}.
		\]
		where the expected frequency $e_{s}$ is computed from the likelihood as the product of the corresponding marginal probabilities multiplied by the sample size.
		
		\item[Step 3d:] Repeat Steps 3b and 3c for $k = (1,\ldots, K)$, for large $K$. Figure~\ref{fig:example} depicts the resulting distribution of $K=1000$ values of the statistic based on 1000 replicated data.
	\end{description}
	
	\item[Step 4:] Compute the proportion of replicated data sets where the value for the statistic was greater than or equal to the value for the observed statistic. 		
	\[
	\text{p}(X^2) = K^{-1} \sum_{k=1}^{K} I(X^{2, (k)}  \geq X^2) 	
	\]
	where the indicator function $I(\cdot)$ equals 1 if the value of the statistic in the replicate data is larger than, or equal to the value of the statistic in the observed data.  
\end{description}

In summary, the goal of the methodology is to obtain the sampling distribution of the statistic of interest using a large number of replicated data sets for which the statistic of interest is computed. This results in $K$ values for the statistic, which are then compared to the value of the statistic in the observed data. As an example Figure \ref{fig:example} displays the values for the Pearson's $X^2$ calculated on $K=500$ data sets generated from a two-class model which was estimated on an empirical data set (which will be discusses later on). The $p$ value is equal to the proportion of generated data sets in which the value for the $X^2$ was at least as large as the observed $X^2$. Visually, this can be represented as the vertical line in Figure~\ref{fig:example} at the value of $X^2 = 226.236$. The $p$ value can then be seen as the area under the curve to the right side of that line. In this example, the $p$ value was equal to 0.266.

Note that we can specify multiple statistics simultaneously in Steps 1 and 3c of the Algorithm. It is important to realise that the implication of a small $p$ value depends on the corresponding statistic. For example, a small $p$ value for the $X^2$ means that the variables in the model-generated data contain less association than in the observed data (only few $X^{2, (k)}$ values were as extreme as the observed $X^2$). The conclusion is then that the given model does not appropriately explain the associations in the observed data.  However, other statistics may have a reverse interpretation, where high $p$ values indicate model misfit. For instance, it may be crucial for an application that a model does neither over- nor underestimate the occurrence of an important response pattern. In such a case, both a high and a low $p$ value for the frequency $n_s$ indicates model misfit.

\subsection{Simulation study.}
We set up a Monte Carlo study to assess type I error rates and power of a number of statistics for evaluating the fit of a LC model. For this, we generated data on $J=6$ dichotomous variables from latent class models with either $C=2$ or $C=3$ classes. Each class was of equal size (i.e. $\rho_c = 1/C,$ for all $c$). Conditional probabilities for a 1 response were $\pi_{jc} =.8$ (or .9) in class 1 and $\pi_{jc} = .2$ (or .1) in class 2 for all $J=6$ dichotomous variables. For class 3, the conditional probabilities for the 1 response were $.8$ (.9) for the first three variables and $.2$ (.1) for the last three variables. The parameter values for the conditions for a three-class model with $\pi_{j1}=.8$ are shown in Table~\ref{tab:lcaparams}. We used sample sizes of $N=100,500$ and 1000. For each combination of conditions we generated 1000 data sets.

\begin{table}[t]
	\centering
	\caption{Class proportions $\classp$ and conditional response probabilities $\condrespone$ for the simulation conditions with $C=3$ classes.}
	\begin{tabular}{cccc}
		\hline
		class           & $c=1$ & $c=2$ & $c=3$\\
		\hline
		$\classp$ & $1/3$  & $1/3$ & $1/3$ \\
		$\pi_{1c}$ & .8 & .2 & .8 \\
		$\pi_{2c}$ & .8 & .2 & .8 \\
		$\pi_{3c}$ & .8 & .2 & .8 \\
		$\pi_{4c}$ & .8 & .2 & .2 \\
		$\pi_{5c}$ & .8 & .2 & .2 \\
		$\pi_{6c}$ & .8 & .2 & .2 \\
		
		\hline
	\end{tabular}
	\label{tab:lcaparams}
\end{table}

For each generated data set we evaluated whether the LC model with 2 classes  was able to reproduce the total association between all variables (quantified by the Pearson's $\chi^2$ and the likelihood ratio $G^2$), the bivariate association between the first two variables (quantified as $\chi^2_{12}$), and the occurrence of the highest-risk pattern with all variables scored 1 (quantified as $\Drisksix$). Data was generated using R 3.3 \citep{r3.3} and subsequent analyses were done with LatentGOLD 5.1 \citep{vermunt2016technical}. The results for the type I error rates and power can be found in Table~\ref{tab:lcares2}.

The first 6 rows of Table~\ref{tab:lcares2} show the type I error rates, which are all very low. That is, regardless of the chosen statistic we rarely rejected the 2-class model as being the data generating process when this was indeed the case. From the last 6 rows, we can see how often we rejected the 2-class model when the 'observed' data was generated from a 3-class model. The power to reject a 2-class model based on aggregated associations in the data was very high (columns 4 and 5),  except when sample size was small ($N=100$) and response probabilities corresponded to $\pi_j1 = .8$. The power to reject the three-class model based on a single bivariate association (column  6) increased overall with bigger sample sizes, but for larger samples, more extreme response probabilities provided lower power. The risk statistic had a power of over .5, except for the smallest sample size and response probabilities of .8.

\begin{table}[t]
	\centering
	\caption{Estimated type-I error rates (rows where $C=2$) and power (rows where $C=3$) when testing the appropriateness of a 2-class LC model when using a significance level of .05. Risk(6) indicates risk statistic for the pattern with all risks present.}
	\begin{tabular}{ccccccc}
		\hline
		$C$   & $N$   & $\pi_{j1}$ & $\chi^2$   & $G^2$ & $\chi^2_{12}$ & $Risk(6)$ \\
		\hline
		2     & 100   & .8    & $.002\pm.002$ & $.006\pm.003$ & $.022\pm.007$ & $.018\pm.006$ \\
		&       & .9    & $.014\pm.005$ & $.006\pm.003$ & $.012\pm.005$ & $.008\pm.004$ \\
		& 500   & .8    & $.000\pm.000$ & $.000\pm.000$ & $.002\pm.002$ & $.002\pm.002$ \\
		&       & .9    & $.000\pm.000$ & $.000\pm.000$ & $.004\pm.003$ & $.000\pm.000$ \\
		& 1000  & .8    & $.000\pm.000$ & $.002\pm.002$ & $.006\pm.003$ & $.002\pm.002$ \\
		&       & .9    & $.000\pm.000$ & $.000\pm.000$ & $.000\pm.000$ & $.002\pm.002$ \\
		3     & 100   & .8    & $.180\pm.017$ & $.214\pm.018$ & $.292\pm.020$ & $.234\pm.019$ \\
		&       & .9    & $.990\pm.004$ & $.956\pm.009$ & $.556\pm.022$ & $.544\pm.022$ \\
		& 500   & .8    & $.934\pm.011$ & $.906\pm.013$ & $.648\pm.021$ & $.500\pm.022$ \\
		&       & .9    & $1.000\pm.000$ & $1.000\pm.000$ & $.544\pm.022$ & $.506\pm.022$ \\
		& 1000  & .8    & $1.000\pm.000$ & $.993\pm.004$ & $.807\pm.018$ & $.593\pm.022$ \\
		&       & .9    & $1.000\pm.000$ & $1.000\pm.000$ & $.659\pm.021$ & $.545\pm.022$ \\
		
		\hline
	\end{tabular}
	\label{tab:lcares2}
\end{table}

\subsection{Application to Empirical Data}

Data from \cite{rindskopf1986value} was used for an empirical illustration. It contains $J=4$ dichotomous variables indicating risk factors of myocardial infarction (MI); that is, presence of Qwave in ECG, presence of flipped LDH, presence of CPK-MB, and presence of classical clinical history. Characteristics of the data were assessed by overall $X^2$ and $G^2$ statistics and bivariate $X^2$s. All total scores were evaluated as well (i.e., Risk statistic with 1, 2, 3, or 4 risks present).

We estimated a 1-class model and a 2-class model, and used our methodology to test the fit of both models. We also compared our methodology with the standard parametric bootstrap. In both methods, the resulting $p$ values were calculated based on $K=1000$ replicated data sets. The results in Table~\ref{tab:empres} indicate that the 1-class model does not capture the associations between the variables nor reproduces the observed frequencies of the risks. All $p$ values were 0 in our method as well as in the parametric bootstrap. For the 2-class model, we did not find any small $p$ value, indicating that the model fits the data well. Our method provided the same conclusion as the computationally more intensive parametric bootstrap.

\begin{table}[h]
	\centering
	\caption{Results of the LC Model Fit Test for the Myocardial Infarction Data.}
	\begin{tabular}{rrrrrrr}
		\hline
		&       & \multicolumn{2}{c}{1-class} &       & \multicolumn{2}{c}{2-class} \\
		Statistic & Value & $p$ value & Bootstrap &       & $p$ value & Bootstrap \\
		\hline
		$X^2$ & 226.236 & .000  & .000  &       & .266  & .308 \\
		$G^2$ & 149.468 & .000  & .000  &       & .490  & .381 \\
		$X^2_{12}$ & 44.082 & .000  & .000  &       & .354  & .213 \\
		$X^2_{13}$ & 39.339 & .000  & .000  &       & .482  & 1.00 \\
		$X^2_{14}$ & 25.034 & .000  & .000  &       & .472  & .288 \\
		$X^2_{23}$ & 41.534 & .000  & .000  &       & .323  & .379 \\
		$X^2_{24}$ & 24.425 & .000  & .000  &       & .379  & .584 \\
		$X^2_{34}$ & 25.824 & .000  & .000  &       & .290  & .225 \\
		Risk(1) & 61    & .000  & NA    &       & .543  & NA \\
		Risk(2) & 46    & .000  & NA    &       & .367  & NA \\
		Risk(3) & 36    & .000  & NA    &       & .633  & NA \\
		Risk(4) & 24    & .000  & NA    &       & .231  & NA \\
		\hline
	\end{tabular}
	\label{tab:empres}
\end{table}

The substantive interpretation of parameters of the 2-class model is straightforward. We encountered a class with low chance of all 4 risks (Class 1 is the group without MI) and a high risk class (Class 2 is the group with MI).

\begin{table}[h]
	\centering
	\caption{Estimated parameter values of a two-class model for the myocardial infarction data. The values for $\pi_{jc}$ give the conditional probabilities of a risk factor being present in that class.}
	\begin{tabular}{rrrr}
		\hline
		&       & Class 1  & Class 2 \\
		\hline
		& $\rho_c$ & .542 & .458 \\
		Q-wave & $\pi_1c$ & .000  & .767 \\
		LDH   & $\pi_2c$ & .027 & .828 \\
		CPK   & $\pi_3c$ & .195 & 1.000 \\
		History & $\pi_4c$ & .195 & .791 \\
		\hline
	\end{tabular}
	\label{tab:emppar}
\end{table}

\section{Discussion}

In this paper we proposed a resampling scheme which combines the ideas of the Bayesian posterior predictive check \citep{meng1994posterior,gelman1996posterior} with frequentist testing procedures, leading to a highly flexible and very fast test for statistical model fit. A detailed description of the methodology was given in the form of a stepwise algorithm. After that the methodology was applied in the context of latent class analysis, where type I error rates and power for different types of model fit tests were evaluated by means of a Monte Carlo study. An application to an empirical data set was provided to test the fit of a LC model used to assess clinical subtypes of patients with risk of myocardial infarction.

The conducted Monte Carlo study showed (very) low type I errors, which follows logically from the methodology. Low type I errors imply that model-generated data was similar to the observed data. This behaviour is also common for the posterior predictive check (e.g., Hjort et al., 2006; van Kollenburg et al., in press). Since the 'observed' data in the simulations was generated from the same model as the replicated data, these are expected to be similar. We found, unsurprisingly, that using a statistic which aggregates all information about the associations in the data had higher power than statistics which used only bivariate associations. The results imply that data generated from a 3-class model has consistently different overall associations than data from a 2-class model. This also holds for bivariate associations, though the power is lower in that case. Interestingly, when sample size was smallest ($N=100$), the bivariate $\chi^2_{12}$ had more power than the aggregated chi-squared statistics. This is likely due to sparseness of the full contingency table. The table had $S=2^6 = 64$ cells with only 100 observations, making calculations of the association measures unreliable. The contingency table for $\chi^2_{12}$ only has 4 cells to fill with 100 observations and sparseness is not an issue. However, it was surprising that at higher sample sizes the response probability had a negative effect on the power of the $\chi^2_{12}$ statistic. Apparently, even with high average cell counts (i.e., when $N/4$ is large) extreme response probabilities may affect the reliability of the statistic. Whether this also holds for bivariate tests in traditional fit testing is an interesting topic for future research.

The results from our simulation can be directly compared to the parametric bootstrap and posterior predictive check by using results from \cite{kollenburg2015assessing}. Those authors compared, among other things, the performance of the parametric bootstrap and the posterior predictive check on type I error rates and power for various statistics in LC analysis. In most cases, the traditional (direct) comparison of model residuals with observed data had higher power. Higher power does come with higher type I error rates as well as significantly longer computation time. Note that in larger sample sizes of $N=1000$ the power of all methods using global fit statistics was approximately 1 already. Future research may focus on comparing different resampling methods more in-depth on type I errors, power, and computation time.

By applying the method to an empirical data set for the standard LC analysis, we found that a 1-class model produced consistently different data sets than the data set we observed, resulting in $p$ values of .000 for all statistics. When we estimated a 2-class model, the data sets that were generated from that model were similar to the observed data. With all $p$ values being between .238 and .538, none of the statistics indicated model misfit. We also performed a parametric bootstrap for this data set, which did not lead to different conclusions about the fit of the model.

All in all, the method proposed in this paper is faster and more flexible than traditional resampling techniques. However, it does provide a more conservative test. There are situations in which conservativeness is a welcome attribute of a test. For instance in substantive research, selecting more classes than necessary in a LC analysis due to random noise in the data (i.e., overfitting) can be more problematic than missing a class \citep{bergh2016building}. Moreover, the current methodology allows researchers to determine exactly which of the aspects they want a model to properly explain/reproduce. As long as the chosen model is able to pick up the characteristics that are of importance to the conducted research, we need not worry about every other aspect of the model fit.  If there are multiple models with which a researcher wants to explain the data, each of these models can be tested with the same methodology.

In this paper we applied our methodology to numerical statistics, and calculated $p$ values to aid us in assessing model fit. In some situations it may also be useful to visualise the data. This is not uncommon in the Bayesian framework. A clear example can be found in \cite{gelman2004bayesian} who even used it as a cover of their book. With minor adjustments to our methodology, such visual tests are now also available in the frequentist setting. The statistic in Step 1 and 3b is then a plot and Step 4 will be a visual inspection rather than the computation of a $p$ value.

On a final note, some researchers require that a test has nominal type I errors. In that case it may be possible to calibrate the resulting $p$ value in much the same way as \cite{kollenburg2017posterior} calibrated the posterior predictive $p$ value. Such a calibration may lead to nominal type I error rates and higher power for the test, but will again require more computation. At which levels of model complexity the computational burden of our method and the parametric bootstrap is the same and what the differences in power are, remain open research questions to answer in the future.

\bibliographystyle{apacite}
\bibliography{AllBibl}

\end{document}